\pdfoutput=1
\documentclass[aps,pre,reprint,longbibliography,floatfix]{revtex4-2}

\usepackage{amsmath,amssymb}
\usepackage{graphicx}
\usepackage[T1]{fontenc}
\usepackage{newtxtext,newtxmath}
\usepackage[hidelinks]{hyperref}
\usepackage{bm}
\usepackage{placeins}
\def \e {\mathrm{e}}

\begin{document}

\title{Controlling synchronization dynamics via physics-informed neural networks}

\author{Kaiming Luo}
\affiliation{School of Information Science and Technology, Fudan University, Shanghai 200438, China}

\date{Received 1 January 2026; accepted 10 June 2026; published 26 June 2026}

\begin{abstract}
Synchronization control in networked dynamical systems requires regulating not only whether coherence is achieved but also when and to what extent it emerges. We propose a physics-informed neural network framework for continuous-time synchronization regulation, in which system trajectories and control inputs are jointly parameterized and constrained by the governing dynamics. Macroscopic synchronization objectives are imposed directly at the trajectory level by enforcing persistence conditions on the order parameter after a prescribed target time. This formulation enables simultaneous control of synchronization time and coherence level without assuming any explicit feedback law or solving a strict optimal control problem. Numerical studies on networked Kuramoto oscillators demonstrate smooth synchronization with reduced transient control effort and competitive cumulative cost relative to analytical baselines. The framework remains effective in nongradient and frustrated dynamics, highlighting physics-informed neural control as a flexible trajectory-level approach to synchronization regulation. This perspective enables controlled access to nontrivial collective states, including chimera and chaotic dynamics.
\end{abstract}

\maketitle

\section{Introduction}

Synchronization is one of the most fundamental collective phenomena in nonlinear dynamics and complex systems, arising ubiquitously in physical, chemical, biological, and engineering contexts \cite{ref01,ref02,ref03,ref04,ref05}. Examples range from Josephson junction arrays \cite{ref06} and laser networks \cite{ref07} to neuronal populations \cite{ref08} and power grids \cite{ref09}, where mutual coupling among individual units leads to the emergence of coherent phase or frequency dynamics. Synchronization concepts have also been extended to hierarchical information dynamics in social media networks, where collective alignment can arise through propagation topology rather than direct interaction \cite{ref10}. The degree of synchronization often plays a decisive role in determining the macroscopic functionality and stability of such systems \cite{ref11}. As a result, the regulation of synchronization processes has long been a central topic in the study of complex dynamical systems \cite{ref12}.

In many practical situations, synchronization is not merely a binary property indicating whether coherence is present or absent, but rather a dynamical process that can, in principle, be regulated \cite{ref13}. On the one hand, it is often desirable for a system to reach a synchronized state within a finite time window, to avoid prolonged desynchronization and the associated degradation of performance or stability \cite{ref14,ref15}. On the other hand, different applications may require different levels of coherence, ranging from partial synchronization to strong phase locking or near-complete synchrony \cite{ref16,ref17,ref18,ref19}. From this perspective, a physically meaningful control problem is to regulate both the timescale over which synchronization emerges and the degree of coherence ultimately attained, subject to the constraints imposed by the underlying dynamics \cite{ref20,ref21}.

A variety of control strategies have been proposed to influence synchronization processes in nonlinear networks, including phase-feedback control, pinning control, edge-based synchronization-suppression strategies, and other structured control schemes \cite{ref22,ref23,ref24}. While these approaches can be effective under specific assumptions, they typically rely on prescribed control laws or local linearization arguments \cite{ref25,ref26}. As a consequence, synchronization performance---including the achieved coherence level, convergence time, and control effort---is usually assessed a posteriori, rather than being specified directly at the trajectory level. This limitation becomes particularly pronounced in high-dimensional, strongly nonlinear, or heterogeneous systems, where shaping the full temporal evolution of collective dynamics remains challenging \cite{ref27,ref28,ref29,ref30}.

In recent years, physics-informed neural networks (PINNs) have emerged as a powerful approach for learning and controlling dynamical systems under physical constraints \cite{ref31,ref32,ref33}. By embedding governing equations directly into the training process of neural networks, PINNs ensure that learned trajectories remain consistent with known physical laws \cite{ref34}. This property makes PINNs particularly attractive for continuous-time problems, where explicit time discretization may introduce numerical stiffness or limit flexibility in trajectory design \cite{ref35}. PINNs have been successfully applied to a wide range of problems, including system identification, inverse problems, and constrained trajectory learning \cite{ref31,ref36}.

Building on these developments, PINNs offer a natural framework for synchronization regulation, in which both system trajectories and control inputs can be represented in a unified, physics-constrained manner \cite{ref33,ref36}. By incorporating system dynamics, initial conditions, and synchronization objectives into a single continuous-time optimization problem, collective observables such as order parameters can be regulated directly at the trajectory level \cite{ref37}. Importantly, this perspective shifts the focus away from prescribing explicit feedback laws or solving optimal control problems in a strict sense, and instead toward discovering admissible and energy-efficient control trajectories that satisfy physical constraints while achieving desired collective behaviors \cite{ref37}.

Motivated by these considerations, we develop a physics-informed neural control framework for the continuous-time regulation of synchronization in networked dynamical systems. The approach reformulates synchronization control as a trajectory-level optimization problem, in which system dynamics, macroscopic synchronization requirements, and control regularization are enforced simultaneously. The governing equations are incorporated through physics-informed residuals, ensuring that the learned trajectories remain dynamically consistent, while synchronization objectives and control cost are imposed directly at the trajectory level. This formulation allows both the synchronization level and the target time to be specified explicitly, without assuming any predefined feedback structure, and recasts synchronization control as the selection of physically admissible trajectories in phase space.

Using networks of coupled oscillators as a representative testbed, we analyze how control effort depends on intrinsic parameters and synchronization targets, and benchmark the learned control against analytically tractable cases such as the Kuramoto model. The framework remains effective in nongradient and frustrated settings where conventional analytical strategies fail. Beyond its control capability, it provides a trajectory-level perspective on how collective states are organized and stabilized, offering a potential probe of the dynamical mechanisms underlying complex synchronization phenomena.

\section{Physics-Informed Neural Control Framework}
\label{sec:framework}

We consider a general class of continuous-time controlled dynamical systems governed by
\begin{equation}
\dot{\mathbf{x}}(t)
=
\mathbf{f}\!\left(\mathbf{p}, \mathbf{x}(t), \mathbf{u}(t)\right),
\label{eq:general_dynamics}
\end{equation}
where $\mathbf{x}(t)\in\mathbb{R}^{d}$ denotes the system state at time $t$, $\mathbf{u}(t)\in\mathbb{R}^{m}$ represents an external control input, and $\mathbf{p}$ is a collection of fixed system parameters, such as coupling strengths or network structure. The nonlinear vector field $\mathbf{f}$ encodes the intrinsic dynamics of the system as well as the manner in which control inputs enter the evolution. The objective of synchronization regulation is not merely to determine whether coherent behavior emerges, but to shape the temporal evolution of collective dynamics generated by Eq.~\eqref{eq:general_dynamics}.

\begin{figure}
    \centering
    \includegraphics[width=1\linewidth]{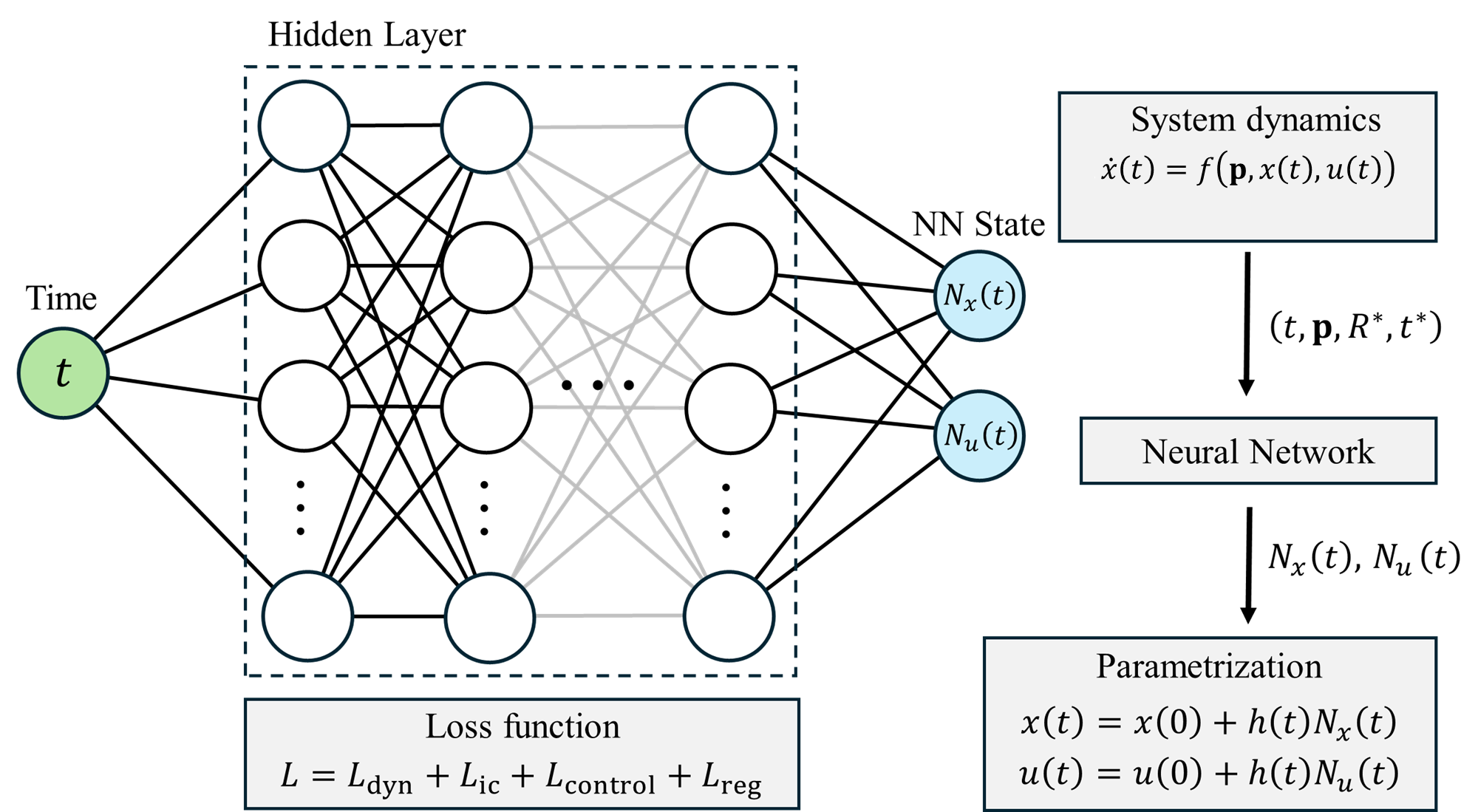}
    \caption{Schematic illustration of the physics-informed neural control framework.
    Time $t$ is used as the input to a neural network with multiple hidden layers, which jointly outputs a state network $N_x(t)$ and a control network $N_u(t)$.
    These outputs are combined with a shaping function $h(t)$ to parametrize the system state $\mathbf{x}(t)$ and control input $\mathbf{u}(t)$ while exactly enforcing the initial condition.
    The governing system dynamics $\dot{\mathbf{x}}(t)=\mathbf{f}(\mathbf{p},\mathbf{x}(t),\mathbf{u}(t))$ are incorporated through physics-informed residuals, and training is performed by minimizing a composite loss function consisting of dynamical, initial-condition, control, and regularization terms.}
    \label{fig:framework}
\end{figure}

The proposed framework represents both the state trajectory $\mathbf{x}(t)$ and the control input $\mathbf{u}(t)$ using neural networks, while enforcing the governing dynamics as physics-informed constraints. As illustrated in Fig.~\ref{fig:framework}, time $t$ is used as the sole explicit input to a feedforward neural network with multiple hidden layers. The network produces two outputs, denoted by $N_x(t)$ and $N_u(t)$, which parametrize the state evolution and control signal, respectively. This joint representation enables the system trajectory and the control input to be learned simultaneously within a unified architecture.

To ensure exact satisfaction of the initial conditions for both the system state and the control input, the neural network outputs are incorporated through the parametrization
\begin{equation}
\begin{aligned}
    &\mathbf{x}(t) = \mathbf{x}(0) + h(t)\, N_x(t),\\
&\mathbf{u}(t) = \mathbf{u}(0) + h(t)\, N_u(t),
\end{aligned}
\label{eq:parametrization}
\end{equation}
where $\mathbf{x}(0)$ denotes the prescribed initial state and $\mathbf{u}(0)$ denotes the initial control value. The function $h(t)=1-\e^{-t}$ is a smooth shaping function satisfying $h(0)=0$, which guarantees that both $\mathbf{x}(t)$ and $\mathbf{u}(t)$ exactly satisfy their respective initial conditions at $t=0$. This construction, shown schematically in Fig.~\ref{fig:framework}, allows the neural networks to flexibly represent admissible state trajectories and control profiles for $t>0$, while preserving consistency with the specified initial conditions.

Consistency between the neural-network parameterized trajectories and the underlying dynamics is enforced by penalizing the residual of Eq.~\eqref{eq:general_dynamics}. Specifically, the physics-informed residual is defined as
\begin{equation}
\mathbf{r}(t)
=
\frac{d\mathbf{x}(t)}{dt}
-
\mathbf{f}\!\left(\mathbf{p}, \mathbf{x}(t), \mathbf{u}(t)\right),
\label{eq:residual}
\end{equation}
where the time derivative $d\mathbf{x}(t)/dt$ is computed via automatic differentiation of the neural network output. Evaluating and penalizing the residual $\mathbf{r}(t)$ at a set of collocation points in time constrains the learned trajectories to remain consistent with the prescribed system dynamics, as conceptually indicated by the feedback between the neural network and the dynamical equation in Fig.~\ref{fig:framework}.
Training of the neural networks is carried out by minimizing a composite loss function of the form
\begin{equation}
\mathcal{L}
=
\mathcal{L}_{\mathrm{dyn}}
+
\mathcal{L}_{\mathrm{control}}
+
\mathcal{L}_{\mathrm{reg}},
\label{eq:loss}
\end{equation}
where $\mathcal{L}_{\mathrm{dyn}}$ penalizes the mean-squared physics-informed residual defined in Eq.~\eqref{eq:residual}, $\mathcal{L}_{\mathrm{control}}$ encodes the synchronization objective, and $\mathcal{L}_{\mathrm{reg}}$ represents the cost associated with the control input. The term $\mathcal{L}_{\mathrm{reg}}$ serves as a regularization that suppresses excessively large control amplitudes and promotes smooth control trajectories, without defining the primary control objective. As indicated in Fig.~\ref{fig:framework}, these loss components jointly guide the training process, enabling the neural network to learn trajectories and controls that satisfy physical constraints while achieving the desired collective behavior.

The framework described above is independent of any specific dynamical model and applies broadly to nonlinear systems exhibiting collective dynamics. In the following sections, networked Kuramoto oscillators are adopted as a representative testbed, owing to their well-defined synchronization order parameter and their role as a standard benchmark for collective behavior. Importantly, the Kuramoto system serves only to instantiate the general physics-informed neural control framework introduced here, which is not restricted to this particular model.

\section{Instantiation on Networked Kuramoto Oscillators}
\label{sec:kuramoto}

To illustrate the general physics-informed neural control framework introduced in Sec.~\ref{sec:framework}, we consider a network of Kuramoto oscillators as a representative testbed. Owing to its simple phase dynamics and well-defined macroscopic order parameter, the Kuramoto model provides a canonical setting for investigating collective synchronization phenomena. Here, however, the Kuramoto system is employed solely to instantiate the general framework, rather than to delimit its applicability.

The dynamics of a network of $N$ coupled phase oscillators are governed by
\begin{equation}
\dot{\theta}_i(t)
=
\omega_i
+
K \sum_{j=1}^{N} A_{ij}
\sin\!\left(\theta_j(t)-\theta_i(t)\right)
+
u_i(t),
\label{eq:kuramoto_control}
\end{equation}
where $\theta_i(t)\in\mathbb{R}$ denotes the phase of oscillator $i$ at time $t$, $\omega_i$ is its intrinsic natural frequency, $K$ is the global coupling strength, and $A_{ij}$ are the elements of the network adjacency matrix. The control input $u_i(t)$ enters additively into the phase dynamics and is learned within the physics-informed neural control framework. The collection of phases $\boldsymbol{\theta}(t)=(\theta_1(t),\ldots,\theta_N(t))$ constitutes the system state appearing in Eq.~\eqref{eq:general_dynamics}.

Synchronization is characterized by the complex order parameter
\begin{equation}
R(t) e^{\mathrm{i}\psi(t)}
=
\frac{1}{N}
\sum_{j=1}^{N}
e^{\mathrm{i}\theta_j(t)},
\label{eq:order_parameter}
\end{equation}
where $R(t)\in[0,1]$ quantifies the instantaneous degree of phase coherence and $\psi(t)$ denotes the mean phase. Incoherent states correspond to values of $R(t)$ close to zero, whereas $R(t)$ approaching unity indicates strong synchronization. Within the present framework, the order parameter $R(t)$ serves as a macroscopic observable through which synchronization objectives are formulated.

Before introducing synchronization control, we verify that the physics-informed neural network can faithfully represent the uncontrolled system dynamics across distinct dynamical regimes. A detailed validation of the PINN trajectory approximation, including both incoherent and spontaneously synchronized regimes of the Kuramoto model, is provided in Appendix~\ref{app:A}. 

For a prescribed synchronization threshold $R^{\ast}\in(0,1)$, the synchronization time is defined \emph{a posteriori} as the trajectory-based quantity
\begin{equation}
T_{\mathrm{s}}(R^{\ast})
=
\inf\Bigl\{
t \ge 0 \;\big|\;
R(t') \ge R^{\ast}, \ \forall\, t' \ge t
\Bigr\},
\label{eq:synchronization_time}
\end{equation}
namely, the earliest time after which the order parameter remains above the threshold $R^{\ast}$. This definition excludes transient crossings and provides a robust characterization of synchronization onset in continuous time.

In the control formulation adopted here, synchronization requirements are specified instead through a \emph{target time} $t^\ast$, which serves as a design parameter rather than a dynamical observable. The control objective is to enforce the inequality
\begin{equation}
T_{\mathrm{s}}(R^\ast) \le t^\ast,
\end{equation}
that is, to require the system to have reached and maintained the prescribed synchronization level by the target time $t^\ast$. The synchronization time $T_{\mathrm{s}}(R^\ast)$ itself is not directly optimized but is constrained implicitly through this requirement.

Within the physics-informed neural control framework, the phase trajectories $\theta_i(t)$ and control inputs $u_i(t)$ are parameterized using neural networks according to Eq.~\eqref{eq:parametrization}, while the Kuramoto dynamics in Eq.~\eqref{eq:kuramoto_control} are enforced through the physics-informed residual defined in Eq.~\eqref{eq:residual}. Synchronization objectives are incorporated by imposing a persistence constraint on the order parameter, requiring $R(t)\ge R^\ast$ for all $t\ge t^\ast$. In this way, both the synchronization level and the time at which it must be achieved are specified explicitly, without prescribing any explicit feedback control law.

For the Kuramoto testbed, the composite loss function introduced in Eq.~\eqref{eq:loss} can be specified explicitly in terms of the phase variables and the synchronization order parameter. In the following loss terms, $w_{\mathrm{dyn}}$, $w_{\mathrm{control}}$, and $w_{\mathrm{reg}}$ are nonnegative weights that tune the relative contributions of dynamical consistency, synchronization enforcement, and control regularization during training. The dynamical loss enforcing consistency with Eq.~\eqref{eq:kuramoto_control} is defined as
\begin{equation}
\begin{aligned}
\mathcal{L}_{\mathrm{dyn}}
= w_{\mathrm{dyn}}
&\frac{1}{N}
\sum_{i=1}^{N}
\Biggl\langle
\Biggl|
\frac{d\theta_i(t)}{dt}
-
\omega_i
\\&-
K \sum_{j=1}^{N} A_{ij}
\sin\!\bigl(\theta_j(t)-\theta_i(t)\bigr)
-
u_i(t)
\Biggr|^{2}
\Biggr\rangle_{t},
\end{aligned}
\label{eq:loss_dyn_kuramoto}
\end{equation}
where $\langle \cdot \rangle_{t}$ denotes averaging over a set of collocation points in time.

The synchronization objective is imposed through a control loss constructed from the order parameter in Eq.~\eqref{eq:order_parameter}. Given the prescribed threshold $R^{\ast}$ and target time $t^\ast$, the synchronization loss is defined as
\begin{equation}
\mathcal{L}_{\mathrm{control}}
= w_{\mathrm{control}}
\left\langle
\mathbf{1}_{t \ge t^\ast}
\bigl[\max\!\left(0,\, R^{\ast}-R(t)\right)\bigr]^{2}
\right\rangle_{t \in [0,T]},
\label{eq:loss_control_kuramoto}
\end{equation}
where $\mathbf{1}_{t \ge t^\ast}$ denotes the indicator function. This formulation enforces the persistence condition $R(t)\ge R^\ast$ for all $t\ge t^\ast$ and thereby implements the constraint $T_{\mathrm{s}}(R^\ast)\le t^\ast$ within a differentiable, physics-informed optimization framework.

The control effort is regularized through a quadratic penalty on the control inputs,
\begin{equation}
\mathcal{L}_{\mathrm{reg}}
= w_{\mathrm{reg}}
\left\langle
\frac{1}{N}
\sum_{i=1}^{N}
u_i(t)^{2}
\right\rangle_{t},
\label{eq:loss_reg_kuramoto}
\end{equation}
This term suppresses excessive control amplitudes and promotes smooth control trajectories.

We emphasize that $\mathcal{L}_{\mathrm{reg}}$ serves purely as a regularization term in the PINN training procedure and is not interpreted as a physical or performance metric. For reporting and comparison purposes, control effort is instead quantified \emph{a posteriori} using the instantaneous control cost $P(t)$ and its time integral $E$, introduced in Sec.~\ref{sec:Sync}.

The accuracy of the physics-informed neural network in approximating uncontrolled dynamics is validated in Appendix~\ref{app:A}.

In the following section, numerical results are presented to demonstrate the ability of the proposed framework to regulate synchronization dynamics in networked Kuramoto oscillators across a range of system parameters.

\section{Numerical Results}

In this section, we present a series of numerical experiments to systematically validate the effectiveness and flexibility of the proposed physics-informed neural control framework for synchronization regulation. The numerical results are organized around three central aspects. First, we demonstrate that the framework can regulate the synchronization process in continuous time, including both the time required to reach synchronization and the achieved level of synchrony. Second, we investigate how the control cost required to enforce a given synchronization objective depends on intrinsic system parameters. Finally, we examine how the control cost varies with the synchronization targets themselves, namely the prescribed synchronization time and synchronization level.

Unless otherwise stated, all numerical experiments are performed on networked systems consisting of a moderate number of oscillators ($N=10$), using the Kuramoto phase model as a representative testbed. This choice allows the synchronization dynamics and the learned control signals to be clearly resolved and systematically analyzed, while avoiding unnecessary obscuration by high-dimensional state representations. The intrinsic frequencies and initial phases are randomly sampled from prescribed distributions, and the network topology and coupling strength are specified or scanned according to the purpose of each experiment. It is important to emphasize that the Kuramoto model is employed here solely as a concrete instantiation of the proposed framework, owing to its well-defined synchronization order parameter. The physics-informed neural control framework itself does not rely on any specific structural property of the Kuramoto dynamics and can, in principle, be extended to more general networked nonlinear dynamical systems. To assess scalability beyond the moderate system sizes considered in the main text, a representative demonstration at a moderately larger network size is reported in Appendix~\ref{app:D}.

Synchronization is quantified using the order parameter \(R(t)\), and the control objective is specified in terms of a target synchronization level \(R^\ast\) and a target time \(t^\ast\), requiring the system to satisfy \(R(t) \ge R^\ast\) for all \(t \ge t^\ast\). Under this constraint, the physics-informed neural network simultaneously learns the system trajectories and the control inputs, such that the governing dynamics, synchronization requirements, and control cost are all satisfied within a unified optimization framework. In the following, we first present results for a representative baseline setting, and then analyze the dependence of the control cost on system parameters and synchronization targets.

\subsection{Synchronization regulation under physics-informed neural control}
\label{sec:Sync}

In this subsection, we consider a representative baseline setting to illustrate the basic capability of the proposed physics-informed neural control framework for synchronization regulation. The purpose of this baseline experiment is not to exhaust the parameter space, but rather to demonstrate that, for a clearly specified synchronization objective, the framework can simultaneously regulate both the synchronization time and the synchronization level in continuous-time dynamics without prescribing an explicit control law.

Specifically, we consider a fully connected network of \(N=10\) oscillators. The intrinsic frequencies \(\omega_i\) are independently drawn from a uniform distribution over the interval \([-\pi,\pi]\), while the initial phases \(\theta_i(0)\) are sampled uniformly from the interval \([-\pi/2,\pi/2]\). The global coupling strength is fixed at \(K=0.05\). The synchronization objective is chosen to be complete synchronization, corresponding to a target synchronization level $R^\ast = 1$, together with a target time $t^\ast = 2$, such that the system is required to satisfy \(R(t) \ge R^\ast\) for all \(t \ge t^\ast\). In practice, $R^\ast=1$ is interpreted in a numerical sense, and synchronization is considered achieved when $R(t)\ge 0.995$ within a prescribed tolerance. Under this setting, the control input is allowed to act over the entire time interval, and no specific functional form of the control law is imposed a priori.

\begin{figure}
    \centering
    \includegraphics[width=1\linewidth]{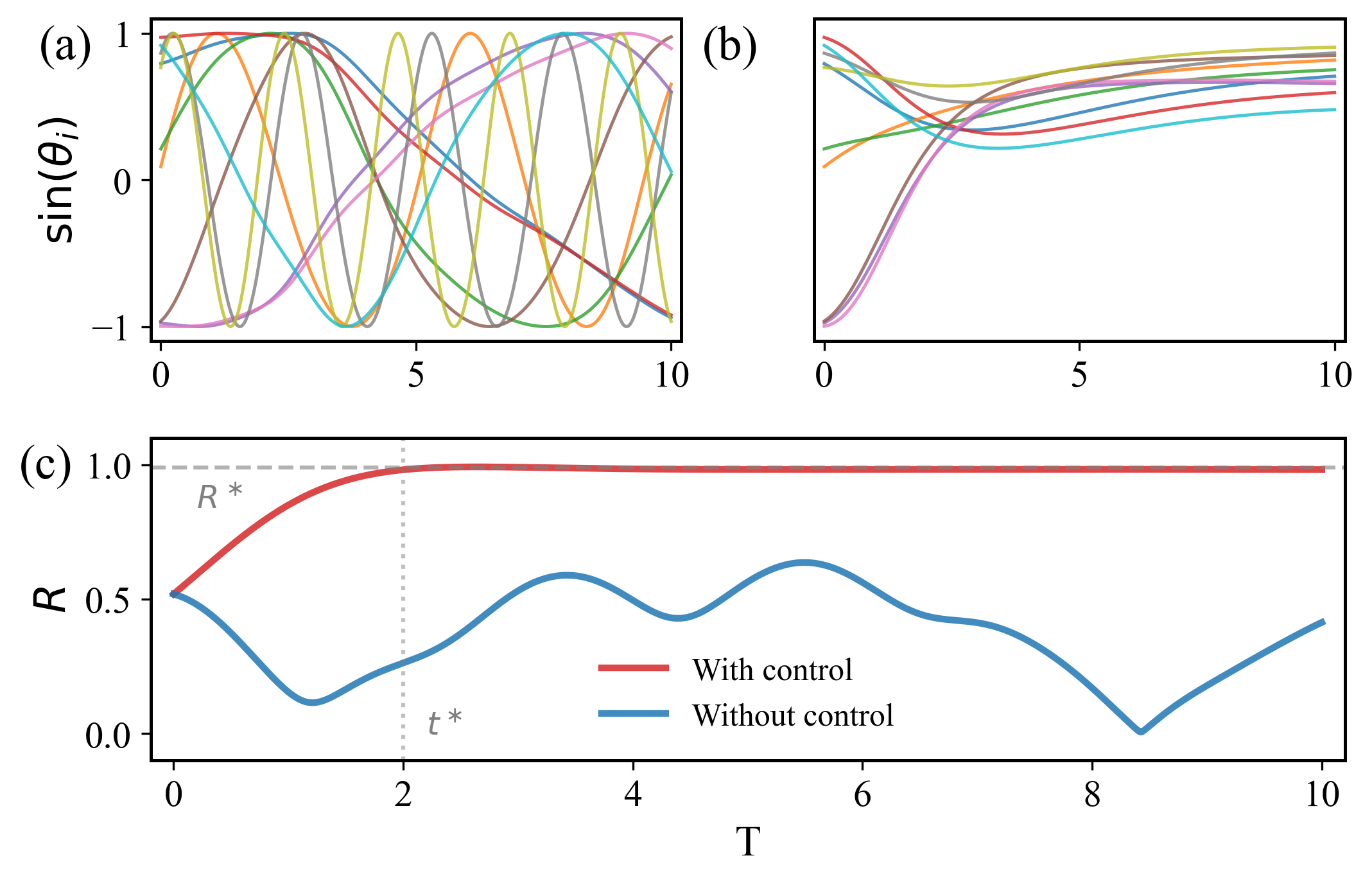}
    \caption{Illustration of synchronization dynamics with and without PINN control. (a) Phase time series of the oscillators without control, represented as $\sin\theta_i(t)$; (b) phase time series under physics-informed neural control; (c) time evolution of the order parameter $R(t)$ for controlled and uncontrolled dynamics. Dashed lines indicate the target synchronization level $R^\ast$ and the target time $t^\ast$.}
    \label{fig:fig2}
\end{figure}

The dynamical evolution of the system in the absence and in the presence of physics-informed neural control is shown in Fig.~\ref{fig:fig2}. Without control, the phase time series of the oscillators, represented as \(\sin\theta_i(t)\), remain dispersed and fail to exhibit coherent collective behavior, as illustrated in Fig.~\ref{fig:fig2}(a). In contrast, under physics-informed neural control, the phase trajectories gradually converge, leading to the formation of a synchronized state within a finite time, as shown in Fig.~\ref{fig:fig2}(b).

This difference is further clarified at the macroscopic level by the evolution of the order parameter. The time dependence of \(R(t)\) for both controlled and uncontrolled dynamics is shown in Fig.~\ref{fig:fig2}(c), together with dashed lines indicating the target synchronization level \(R^\ast\) and the target time \(t^\ast\). In the absence of control, the order parameter remains well below the target level throughout the time interval. By contrast, when control is applied, \(R(t)\) increases rapidly and reaches \(R^\ast=1\) before \(t^\ast\), after which it remains close to unity. These results demonstrate that the proposed framework enables precise regulation of both the synchronization time and the synchronization level in continuous time.

\begin{figure}
    \centering
    \includegraphics[width=1\linewidth]{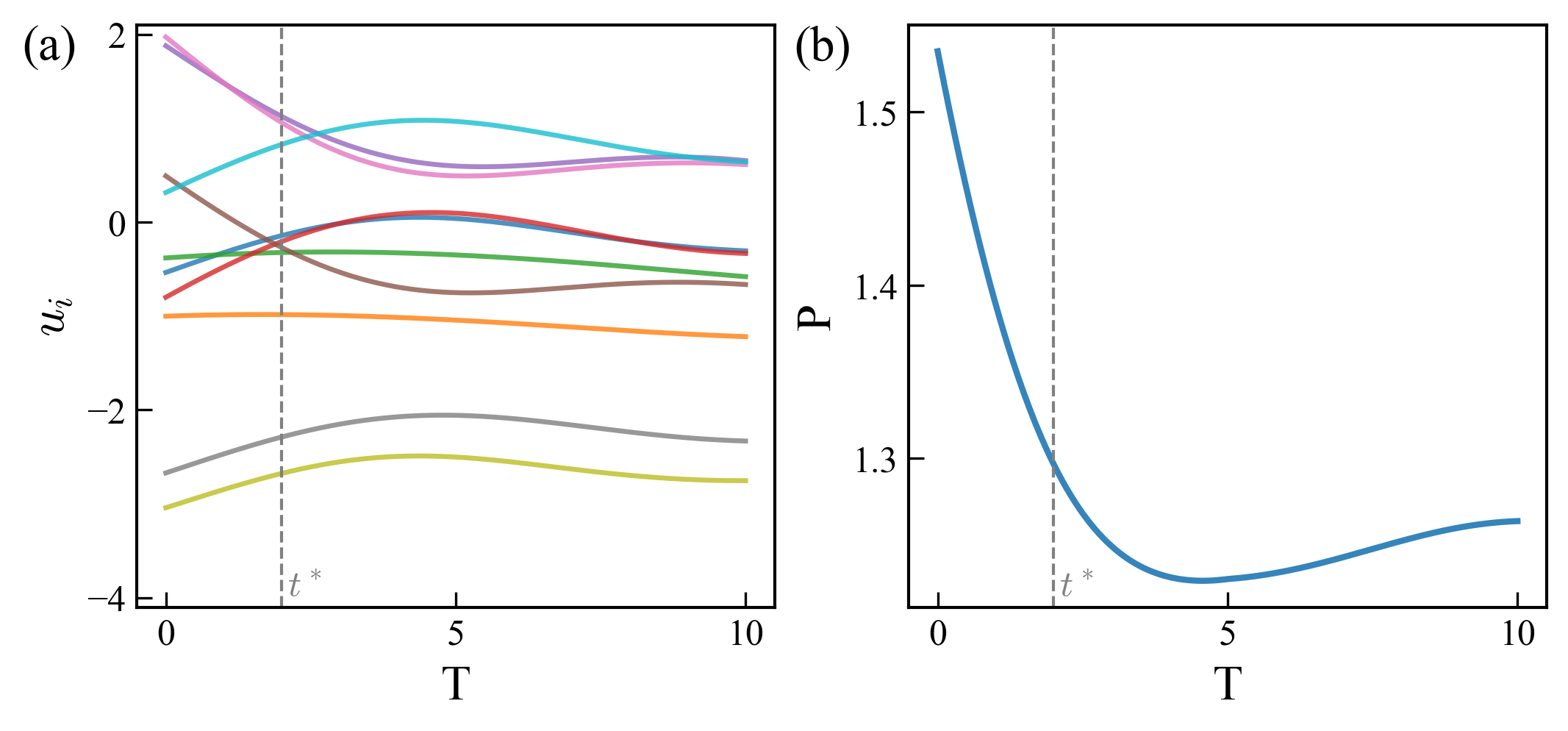}
    \caption{Control signals and associated effort under PINN control. (a) Time evolution of the control inputs $u_i(t)$ learned by the physics-informed neural network; 
(b) control cost density $P$ as a function of time, quantifying the instantaneous control effort required to enforce the synchronization constraint.
}
    \label{fig:fig3}
\end{figure}

Having established that the synchronization objective is successfully achieved, we next examine the structure of the control signals learned by the physics-informed neural network. The time evolution of the control inputs \(u_i(t)\) applied to individual oscillators is shown in Fig.~\ref{fig:fig3}(a). The control signals are continuous and smooth over the entire time horizon, with no evidence of impulsive or high-frequency behavior. As the system approaches the synchronized state, the magnitude of the control inputs decreases, indicating that only weak control is required to maintain synchronization once it has been established.

To quantify the instantaneous control effort, we introduce the instantaneous control cost
\begin{equation}
P(t) = \frac{1}{N}\sum_{i=1}^{N} u_i^2(t),
\label{eq:P(t)}
\end{equation}
which measures the magnitude of the applied control inputs at time $t$.
This quantity is a dimensionless cost measure associated with control amplitudes and should not be interpreted as the physical power or energy of the oscillator network.

As shown in Fig.~\ref{fig:fig3}(b), the control cost is relatively large during the synchronization formation stage and decreases substantially once the synchronized state is established, eventually approaching a steady value. This behavior indicates that the control action primarily facilitates the transient process leading to synchronization, while only a weak and sustained control effort is required to maintain the synchronized state thereafter.

To further illustrate this model independence, we apply the same framework to networks of Stuart--Landau oscillators, which include both amplitude and phase degrees of freedom; representative results are reported in Appendix~\ref{app:C}.

\subsection{Control cost under different conditions}

In this subsection, we systematically investigate how the control expenditure required to achieve a prescribed synchronization objective depends on both intrinsic system properties and the specification of the control targets. Throughout this analysis, the network size and the PINN control framework are kept fixed, while different sources of synchronization difficulty are varied in a controlled manner. This allows us to quantitatively assess how distinct mechanisms that hinder synchronization are reflected in the required control cost.

To enable consistent comparisons across different conditions, we define the integrated control cost as
\begin{equation}
E = \int_{0}^{T} P(t)\,\mathrm{d}t,
\label{eq:control_energy}
\end{equation}
where $P(t)$ denotes the instantaneous control cost defined in Eq.~\eqref{eq:P(t)}, and $T$ is the terminal time of the control horizon. Due to the normalization by the network size $N$ embedded in $P(t)$, the quantity $E$ represents the average cumulative control effort per oscillator. This normalization ensures that variations in $E$ can be directly interpreted as changes in synchronization difficulty rather than trivial scaling with system size.

\begin{figure}
    \centering
    \includegraphics[width=1\linewidth]{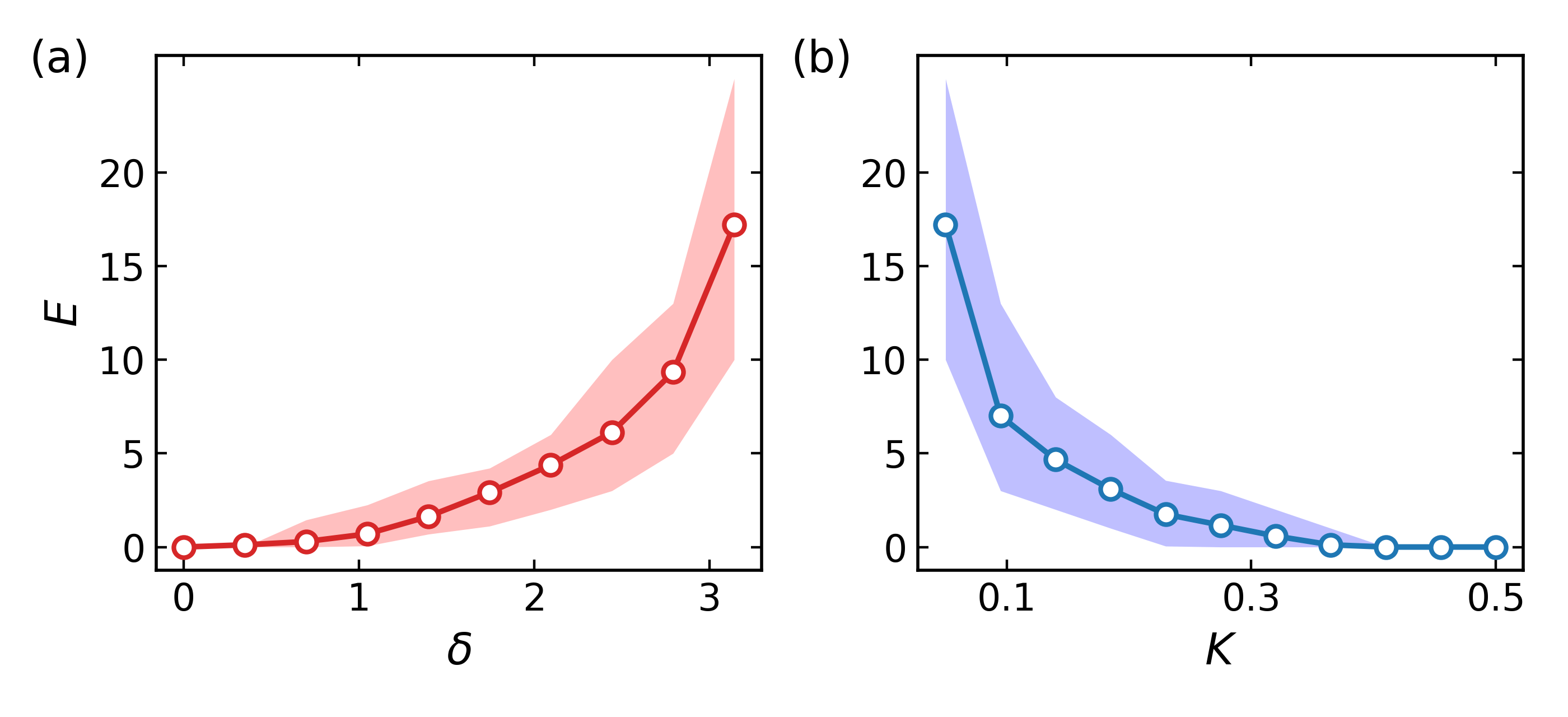}
    \caption{
Integrated control cost $E$ under different intrinsic system conditions.
(a) Dependence of $E$ on the frequency heterogeneity width $\delta$.
(b) Dependence of $E$ on the coupling strength $K$.
In both panels, the synchronization target $(R^\ast, t^\ast)$ and all other system parameters are fixed.
Solid lines indicate mean values, while shaded regions represent variability across independent realizations; for each parameter value, results are averaged over $20$ independent runs with different random initial conditions and frequency realizations.
}
    \label{fig:fig4}
\end{figure}

We first examine the dependence of the integrated control cost on intrinsic system parameters while keeping the synchronization target $(R^\ast, t^\ast)$ fixed, as shown in Fig.~\ref{fig:fig4}. Figure~\ref{fig:fig4}(a) reports the variation of $E$ as a function of the frequency heterogeneity width $\delta$, which characterizes the spread of intrinsic oscillator frequencies. As $\delta$ increases, the intrinsic frequencies become more broadly distributed, enhancing the tendency of the system to desynchronize. To compensate for this increased heterogeneity and enforce the same synchronization target, the controller must supply stronger and/or more persistent inputs, leading to a monotonic increase in the integrated control cost. This trend quantitatively captures the growing intrinsic difficulty of synchronizing a heterogeneous population of oscillators.

An opposite dependence is observed when varying the coupling strength $K$, as shown in Fig.~\ref{fig:fig4}(b). Increasing $K$ enhances the effectiveness of mutual interactions in promoting phase alignment, thereby reducing the reliance on external control. Consequently, the integrated control cost decreases monotonically with increasing coupling strength. From a dynamical perspective, increasing $K$ effectively counteracts frequency heterogeneity by strengthening collective synchronization tendencies. It is therefore natural that variations in $K$ and $\delta$ lead to opposite trends in the control cost. The complementary behavior observed in Fig.~\ref{fig:fig4} further supports the interpretation of $E$ as a quantitative proxy for intrinsic synchronization difficulty.

We next investigate how the integrated control cost depends on the synchronization targets themselves, focusing on the required coherence level and the prescribed synchronization time. The quantity $E$ defined in Eq.~\eqref{eq:control_energy} provides a direct measure of the cumulative effort needed to meet a given control objective.

\begin{figure}
    \centering
    \includegraphics[width=1\linewidth]{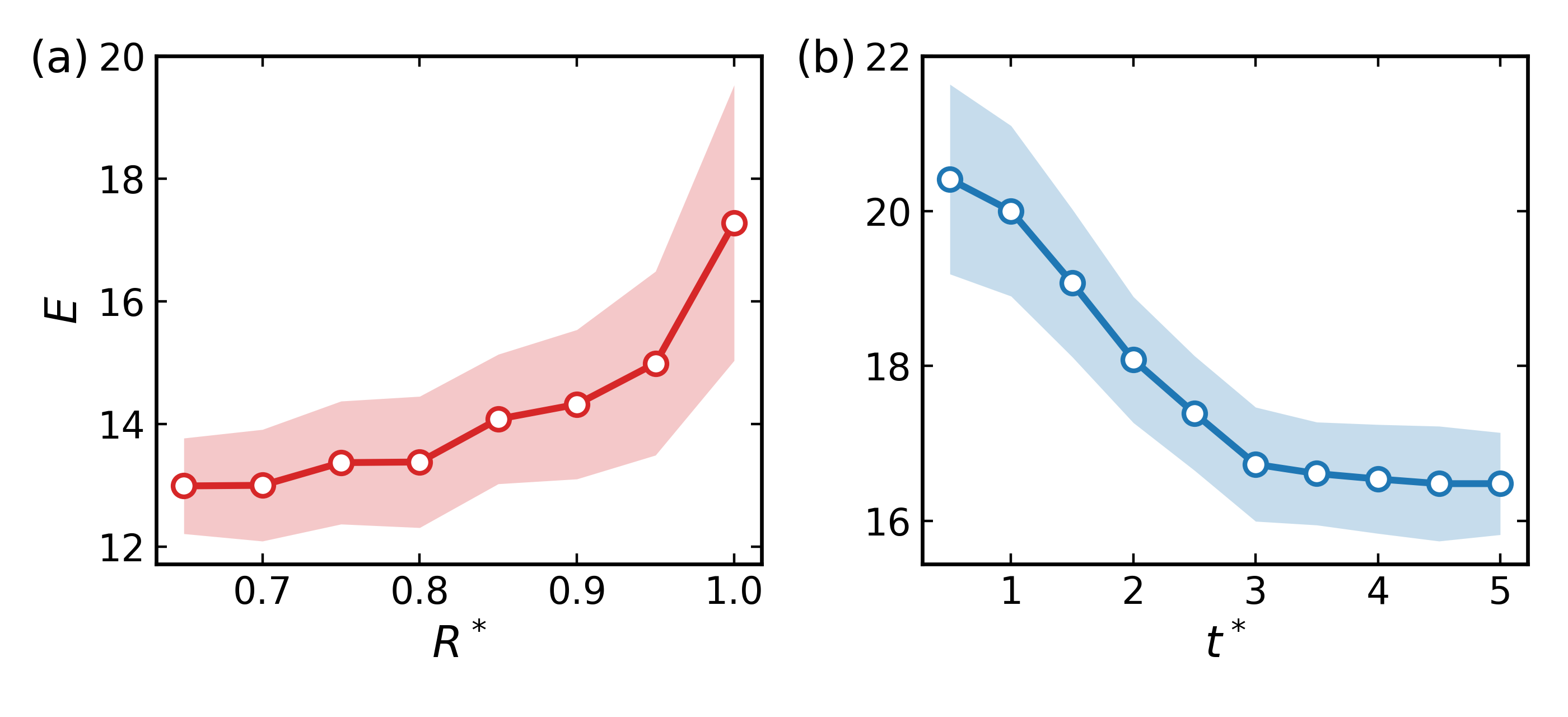}
    \caption{
Dependence of the integrated control cost $E$ on synchronization targets.
(a) Integrated control cost as a function of the target synchronization level $R^\ast$ with fixed target time $t^\ast$.
(b) Integrated control cost as a function of the target synchronization time $t^\ast$ with fixed $R^\ast$.
Solid lines denote mean values over $20$ independent runs, and shaded regions indicate the corresponding variability.
}
    \label{fig:fig5}
\end{figure}

We first consider the dependence of $E$ on the target synchronization level $R^\ast$ while keeping the target time $t^\ast$ fixed, as shown in Fig.~\ref{fig:fig5}(a). As $R^\ast$ increases, the integrated control cost grows monotonically and rises sharply as $R^\ast$ approaches unity. This behavior reflects the increasing difficulty of enforcing near-complete phase coherence: achieving higher synchronization accuracy requires progressively finer phase alignment, which in turn demands stronger and more sustained control actions to suppress residual phase fluctuations.

We then examine the complementary dependence of $E$ on the target synchronization time $t^\ast$, with the synchronization level $R^\ast$ held fixed, as shown in Fig.~\ref{fig:fig5}(b). In this case, the integrated control cost decreases systematically as $t^\ast$ increases. Enforcing synchronization within a shorter time window requires more aggressive control interventions to accelerate convergence toward the synchronized manifold, thereby increasing the cumulative control cost. Allowing a longer synchronization time enables the intrinsic coupling dynamics to contribute more effectively, reducing the burden on external control.

Across the entire parameter range considered, the variability of the integrated control cost over repeated training runs remains moderate, indicating that the observed trends are robust with respect to stochastic effects associated with random initial conditions and neural network optimization.

Overall, these results reveal a clear trade-off between synchronization performance and control expenditure: achieving higher coherence or faster convergence inevitably incurs higher control cost. By explicitly quantifying this trade-off within a continuous-time, physics-informed framework, the proposed approach provides a principled tool for exploring cost--performance landscapes in synchronization regulation problems.

\subsection{Comparison with baseline strategies}

In this subsection, we benchmark the proposed PINN control framework against representative baseline strategies for synchronization regulation, with particular emphasis on control energy efficiency.
Rather than aiming to solve an optimal control problem in a strict theoretical sense, the goal of this comparison is to assess whether the PINN-based framework can systematically identify low-energy admissible control trajectories under identical dynamical constraints and synchronization objectives.

All methods are evaluated under the same system parameters, including identical distributions of initial conditions and intrinsic frequencies across multiple realizations. Synchronization is achieved in all realizations under the reported settings, and all results are averaged over independent runs without selection, ensuring a consistent and fair comparison across methods. The comparison focuses on both transient control behavior, quantified by the instantaneous control cost $P(t)$, and cumulative control expenditure, quantified by the integrated control cost $E$.

The regularization weight $w_{\mathrm{reg}}$ is chosen in the range $10^{-4}$--$10^{-2}$, which ensures accurate satisfaction of the governing dynamics and synchronization constraints while promoting low-energy solutions. Within this range, the qualitative behavior of the learned control trajectories and the relative comparison with baseline strategies remain robust.

We consider three representative baseline strategies commonly used for networked phase oscillators synchronization control: linear phase-based feedback, nonlinear (sinusoidal) phase-based feedback, and frequency-based compensation.
The linear phase-based control is given by
\begin{equation}
u_i(t) = -k_\theta\bigl(\theta_i(t)-\bar{\theta}(t)\bigr),
\end{equation}
while the nonlinear phase-based control takes the form
\begin{equation}
u_i(t) = -k_\theta \sin\bigl(\theta_i(t)-\bar{\theta}(t)\bigr),
\end{equation}
where $\bar{\theta}(t)$ denotes the instantaneous mean phase.
In contrast, the frequency-based strategy compensates intrinsic frequency heterogeneity via
\begin{equation}
u_i(t) = -k_\omega(\omega_i-\bar{\omega}),
\end{equation}
with $\bar{\omega}$ being the mean intrinsic frequency.
For the Kuramoto model, frequency-based compensation admits an analytic solution and thus provides a natural low-energy reference within additive control schemes. In this setting, it serves as a benchmark for assessing whether the PINN, without explicitly incorporating this analytical structure, can recover control trajectories with comparable efficiency. It should therefore be interpreted as a problem-specific reference rather than a general optimality benchmark.

\begin{figure}
    \centering
    \includegraphics[width=\linewidth]{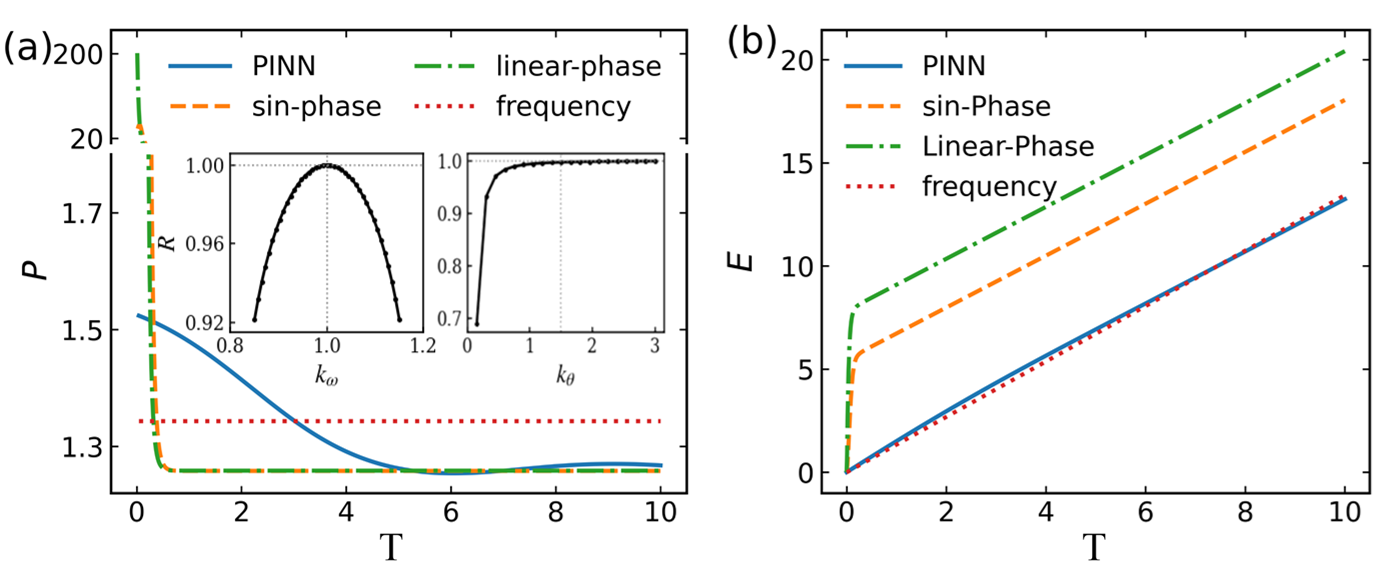}
    \caption{
Benchmark comparison between PINN-based control and baseline strategies in the Kuramoto model.
(a) Instantaneous control cost $P(t)$.
(b) Integrated control cost $E$.
}
    \label{fig:baseline_comparison}
\end{figure}

For each baseline method, the control gain is determined by parameter scanning and selected as the minimal value that satisfies the persistence-based synchronization criterion defined in Eq.~\eqref{eq:synchronization_time}, namely that $R(t)\ge R^\ast$ for all $t\ge t^\ast$. The same criterion is used consistently for all methods.

As indicated by the inset parameter scans in Fig.~\ref{fig:baseline_comparison}(a), phase-based feedback requires a gain of $k_\theta=1.5$, whereas frequency-based compensation achieves synchronization at $k_\omega=1$.
These values are used consistently throughout the comparison to ensure identical synchronization performance and a fair assessment of transient behavior and control cost.

The temporal evolution of the instantaneous control cost $P(t)$ reveals a clear qualitative distinction between the PINN-based control and the phase-based baselines, as shown in Fig.~\ref{fig:baseline_comparison}(a).
Both linear and nonlinear phase feedback exhibit pronounced transient peaks at early times, reflecting their strong sensitivity to the initial phase configuration.
When oscillators are initially far from coherence, large corrective inputs are required to rapidly suppress phase dispersion, resulting in abrupt and intense control actions.
Such transient behavior may pose practical limitations in systems subject to actuator saturation or amplitude constraints.

By contrast, the PINN-based control produces smooth and gradually varying control profiles without sharp transients.
By explicitly incorporating the governing dynamics and synchronization objectives during training, the PINN distributes control effort more evenly over time, thereby biasing the learned trajectories toward energy-efficient admissible solutions rather than aggressively correcting instantaneous phase deviations.

The accumulated control cost further highlights the differences between control strategies.
As shown in Fig.~\ref{fig:baseline_comparison}(b), both phase-based feedback schemes incur substantially larger integrated control cost, dominated by their large early-time transients.
In this analytically tractable setting, frequency-based compensation provides a meaningful low-energy reference.
Remarkably, the cumulative control cost achieved by the PINN is found to be comparable to this analytical reference, despite the absence of any explicitly prescribed frequency-compensation control law.
This observation indicates that the PINN-based framework is capable of discovering highly energy-efficient control trajectories directly from the governing dynamics, without invoking optimal control formulations or analytic solutions.

The above comparison relies on the existence of an analytically solvable frequency-compensation strategy, which is specific to gradient synchronization dynamics such as the Kuramoto model.
In such systems, intrinsic frequency heterogeneity can be compensated analytically, providing a meaningful low-energy reference for additive control schemes.
However, this type of analytical baseline ceases to exist once the underlying gradient structure is broken.

To assess the robustness of the proposed PINN control framework beyond gradient systems and to illustrate the limitations of analytical references, we consider the Kuramoto--Sakaguchi model,
\begin{equation}
\dot{\theta}_i(t) = \omega_i + K\sum_{j=1}^N \sin\!\bigl(\theta_j(t)-\theta_i(t)-\alpha\bigr) + u_i(t),
\end{equation}
where the phase lag $\alpha$ introduces intrinsic frustration.
This phase shift breaks the potential structure underlying the Kuramoto model and renders the dynamics nongradient, thereby eliminating the analytical tractability of frequency-based compensation strategies.

\begin{figure}
    \centering
    \includegraphics[width=\linewidth]{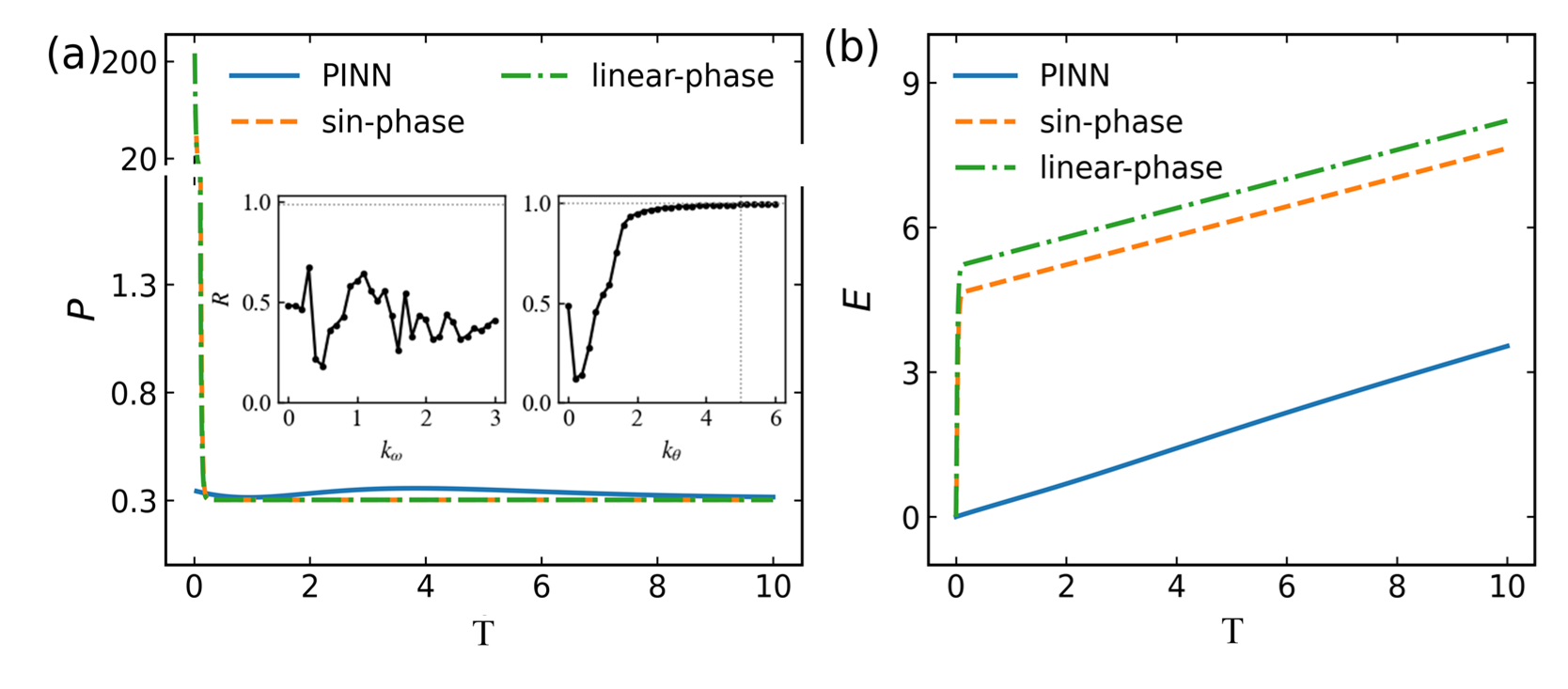}
    \caption{
Benchmark comparison between PINN-based control and baseline strategies in the Kuramoto-Sakaguchi model.
(a) Instantaneous control cost $P(t)$.
(b) Integrated control cost $E$.
}
    \label{fig:KS_comparison}
\end{figure}

The control performance in a representative frustrated regime with $\alpha=\pi/2$ and coupling strength $K=0.4$ is summarized in Fig.~\ref{fig:KS_comparison}.
In this regime, the uncontrolled system does not admit a synchronized state with $R\approx1$, and frequency-based compensation fails to stabilize synchronization, thereby ceasing to provide a meaningful reference.
This breakdown is reflected in both the instantaneous and accumulated control costs shown in Figs.~\ref{fig:KS_comparison}(a) and~\ref{fig:KS_comparison}(b).

Phase-based feedback can still enforce synchronization under these nongradient dynamics.
However, as shown in Fig.~\ref{fig:KS_comparison}(a), this is achieved at the expense of pronounced transient peaks in the instantaneous control cost $P(t)$, arising from the strong sensitivity of phase feedback to instantaneous phase dispersion.
These early-time peaks dominate the accumulated control cost, leading to a substantially larger integrated control cost $E$, as shown in Fig.~\ref{fig:KS_comparison}(b).

By contrast, the PINN-based framework remains effective in this frustrated regime.
The learned control signals yield smooth temporal profiles without sharp transients, resulting in consistently reduced instantaneous control cost and a significantly lower integrated control cost compared to phase-based feedback.
Importantly, this improvement is achieved without relying on any analytic structure specific to gradient systems.

This comparison highlights the methodological advantage of the physics-informed neural control framework.
Rather than exploiting problem-specific analytical properties, the PINN performs a physics-constrained search over admissible control trajectories and is naturally biased toward low-energy solutions through its training objective.
As a result, the framework provides a flexible and energy-efficient control discovery mechanism that remains effective even in frustrated and nongradient synchronization dynamics, where conventional analytical baselines and optimality notions are no longer available.

\subsection{Robustness to noise}
\label{sec:noise}

In practical applications, synchronization control is inevitably affected by stochastic perturbations arising from environmental fluctuations, unmodeled dynamics, or measurement uncertainty. It is therefore essential to assess whether the proposed PINN control strategy can maintain stable synchronization under noisy conditions, despite being designed as an offline trajectory-level controller.

To this end, we examine the robustness of the learned control signals by introducing additive stochastic perturbations into the controlled Kuramoto dynamics during the \emph{execution stage}, while keeping the PINN-trained control inputs fixed. Specifically, after training the PINN under deterministic dynamics, the system is evolved according to the stochastic differential equation
\begin{equation}
\dot{\theta}_i(t)
=
\omega_i
+
K \sum_{j=1}^{N} A_{ij} \sin\!\bigl(\theta_j(t)-\theta_i(t)\bigr)
+
u_i(t)
+
\sigma\,\xi_i(t),
\label{eq:noise_kuramoto}
\end{equation}
where $\xi_i(t)$ denotes independent Gaussian white noise satisfying
$\langle \xi_i(t) \rangle = 0$ and
$\langle \xi_i(t)\xi_j(t') \rangle = \delta_{ij}\delta(t-t')$.
The parameter $\sigma$ controls the noise intensity.
Unless otherwise stated, the control signals $u_i(t)$ are those learned from noiseless PINN training and are not adapted online.

The robustness of the PINN-based control is quantified using three complementary metrics:
(i) the time-averaged order parameter
$\langle R \rangle = (T-t^*)^{-1}\int_{t^*}^{T} R(t)\,dt$,
(ii) the relative synchronization error
$\mathcal{E}_R = (R^{*}-\langle R \rangle)/R^{*}$,
and (iii) the integrated control cost
$E$. For each noise intensity $\sigma$, all quantities are averaged over multiple independent noise realizations.

\begin{figure}
    \centering
    \includegraphics[width=1\linewidth]{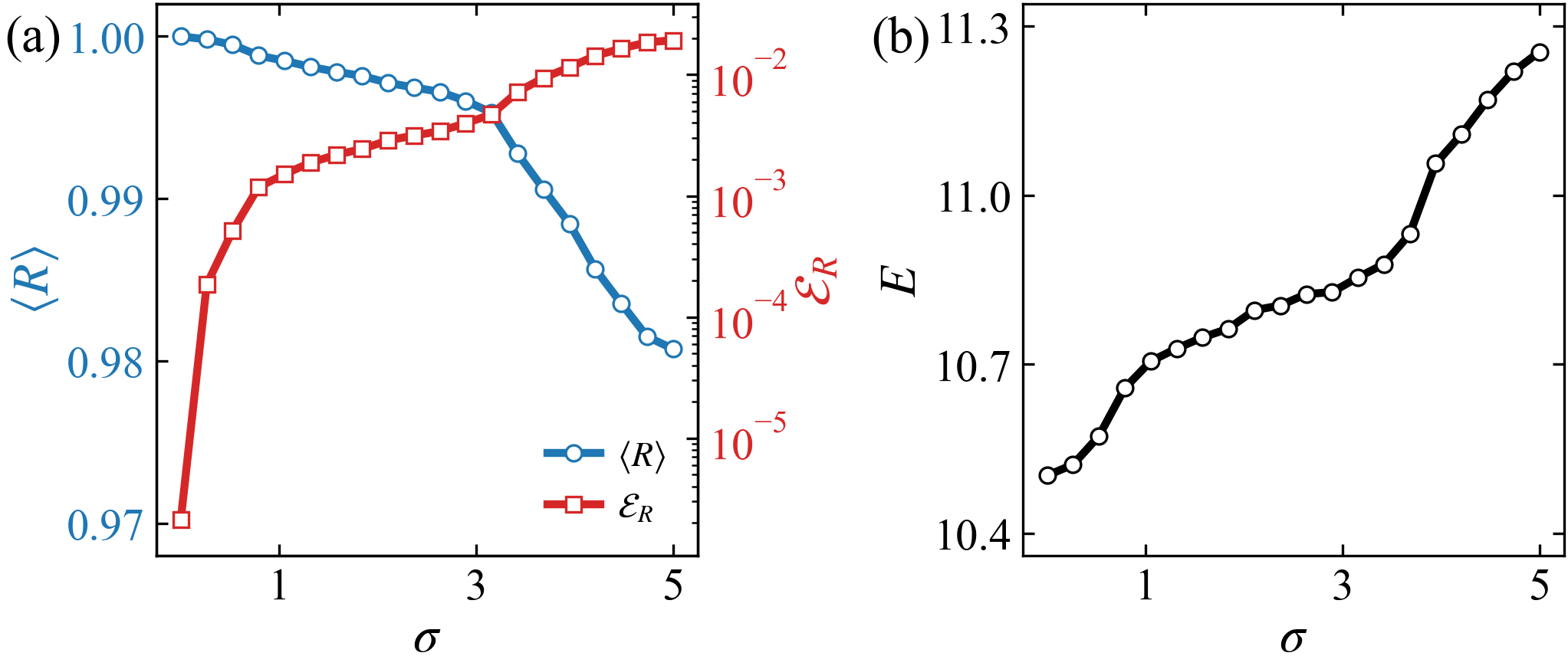}
    \caption{
    Noise robustness analysis of the PINN-controlled system.
    (a) Time-averaged order parameter $\langle R \rangle$ (left axis) and relative synchronization error $\mathcal{E}_R$ (right axis) as functions of the noise intensity $\sigma$ for a fixed coupling strength and target $R^{*}=1$.
    (b) Integrated control cost $E$ as a function of the noise intensity $\sigma$ under the same noisy conditions.
    }
    \label{fig:noise_scan}
\end{figure}

Figure~\ref{fig:noise_scan}(a) shows the dependence of synchronization performance on the noise intensity $\sigma$ for a fixed coupling strength and a target synchronization level $R^{*}=1$. As $\sigma$ increases, the time-averaged order parameter $\langle R \rangle$ decreases gradually from unity, indicating a progressive degradation of phase coherence induced by stochastic perturbations. At the same time, the synchronization error $\mathcal{E}_R$ increases monotonically with $\sigma$, reflecting enhanced fluctuations around the synchronized manifold rather than an abrupt loss of global synchronization. Importantly, synchronization remains robust over a broad range of noise intensities, with $\langle R \rangle$ staying close to unity even for moderately strong noise.

The corresponding variation of the integrated control cost $E$ as a function of the noise intensity $\sigma$ is reported in Fig.~\ref{fig:noise_scan}(b). Because the control inputs are fixed offline and not adjusted in response to instantaneous noise realizations, the overall magnitude of $E$ remains finite and varies smoothly with increasing $\sigma$. The observed increase in $E$ reflects the additional control effort required to suppress noise-induced phase fluctuations and to stabilize the collective dynamics against stronger stochastic perturbations.

Taken together, these results demonstrate that the PINN-based control strategy exhibits inherent robustness to stochastic perturbations. Rather than reacting aggressively to instantaneous noise realizations, the learned control stabilizes synchronization at a global, trajectory level, ensuring that macroscopic coherence is maintained with only a modest and smoothly varying increase in control cost. This robustness, achieved without online feedback or adaptive re-optimization, underscores the practical viability of the proposed physics-informed control framework in noisy environments.

\section{Discussion}

The results presented in this work establish PINNs as an effective and flexible framework for continuous-time synchronization regulation in networked nonlinear dynamical systems. Beyond demonstrating the realization of prescribed synchronization objectives in the Kuramoto testbed, the primary contribution of this study is methodological: synchronization control is reformulated as a unified, physics-constrained, trajectory-level optimization problem, in which admissible system trajectories and control inputs are learned simultaneously while macroscopic synchronization requirements are imposed explicitly in time.

A key conceptual distinction between the proposed PINN-based framework and conventional synchronization control strategies lies in the nature of the learned object. Rather than prescribing an explicit feedback law or control structure \emph{a priori}, the PINN parameterizes entire state and control trajectories over a finite time horizon and enforces dynamical consistency through physics-informed residuals. From this perspective, training the PINN can be interpreted as solving an implicit continuous-time constrained optimization problem: among all trajectories that satisfy the governing equations and synchronization constraints, those minimizing a prescribed control regularization are selected. As a result, the learned control does not correspond to a local-in-time reaction to instantaneous phase deviations, but instead represents a globally optimized intervention profile that distributes control effort coherently across time. This global-in-time character naturally explains the smooth temporal structure of the learned control signals and the absence of impulsive or high-frequency components.

The systematic comparison with phase-based baseline strategies highlights an intrinsic limitation of instantaneous phase feedback. Both linear and nonlinear phase-feedback controls act directly on deviations from the instantaneous mean phase. When the initial phase configuration is highly incoherent, large phase dispersion leads to strong corrective forces at early times, resulting in pronounced transient peaks in the instantaneous control cost. These peaks are not numerical artifacts but arise from the fundamental sensitivity of phase-based feedback to initial conditions and instantaneous phase geometry. In practical scenarios involving actuator saturation, amplitude constraints, or power limitations, such large transient demands may significantly restrict the applicability of purely phase-based control schemes, even when synchronization can eventually be achieved.

The frequency-compensation baseline occupies a distinct conceptual role. For the Kuramoto model, compensating intrinsic frequency heterogeneity constitutes an analytically tractable strategy and provides a meaningful reference for assessing control efficiency in this specific gradient system. The observation that the integrated control cost achieved by the PINN-based approach closely matches that of frequency compensation admits a clear interpretation: by exploiting the governing dynamics together with explicit synchronization constraints during training, the PINN is able to identify a control trajectory that is effectively low-cost under specified conditions for this analytically solvable case, without explicitly encoding the corresponding control law. Importantly, this agreement should be viewed as a validation of the framework in a setting where a theoretical benchmark exists, rather than as evidence of any universal optimality guarantee. For more general nonlinear or nongradient systems, analytic frequency-based controls are typically unavailable, whereas the PINN formulation remains directly applicable.

This distinction becomes particularly relevant in frustrated or nongradient dynamics, such as the Kuramoto--Sakaguchi model examined in this work. In such settings, frequency-based compensation ceases to provide a meaningful baseline, and phase-based feedback often incurs substantially increased transient and cumulative control costs. By contrast, the PINN-based framework continues to produce smooth and effective synchronization control, underscoring that its performance does not rely on special gradient structures of the underlying dynamics. This robustness to the qualitative nature of the dynamical system highlights the promising applicability of the physics-informed trajectory optimization perspective.

The applicability of the framework is further supported by the extensions reported in Appendixes~\ref{app:C} and~\ref{app:D}. In particular, the successful application to networks of Stuart--Landau oscillators demonstrates that the approach naturally extends beyond pure phase models to systems with both amplitude and phase degrees of freedom. This confirms that the framework does not rely on specific properties of the Kuramoto order parameter but instead applies whenever a suitable macroscopic collective observable can be defined. Moreover, representative demonstrations at larger network sizes indicate that the approach remains feasible beyond the small systems used for detailed analysis in the main text. While increasing system size raises representational and optimization demands, these challenges are primarily technical rather than conceptual and can be addressed by increasing network capacity, refining collocation strategies, and leveraging parallel computation.

An additional advantage of the PINN formulation lies in the explicit manner in which synchronization objectives are specified and enforced. In conventional approaches, control laws are designed first and synchronization performance---both in terms of coherence level and convergence time---is assessed \emph{a posteriori}. In contrast, the PINN framework incorporates macroscopic synchronization requirements directly into the continuous-time learning objective. Targets specifying both the desired synchronization level and the time at which it must be reached and maintained are imposed explicitly within the optimization problem, rendering synchronization performance a programmable specification rather than an emergent byproduct of a chosen feedback structure.

This capability also carries direct implications for the exploration and control of nontrivial dynamical regimes in oscillator networks. By embedding macroscopic constraints into the learning objective, the framework enables controlled access to nontrivial collective states---including chimera patterns, cluster synchronization, and chaotic dynamics---that are typically obtained only through indirect parameter tuning. Such states can instead be imposed as dynamical targets with prescribed temporal structure, providing a systematic way to probe partial synchronization, multistability, and instability mechanisms while remaining consistent with the governing dynamics.

Several limitations of the present study should be acknowledged. The current implementation constitutes an offline design procedure: the learned control corresponds to a trajectory optimized for a given objective and set of conditions, rather than a real-time adaptive feedback controller. Training costs increase with system dimension and with the strictness of imposed constraints, and robustness to model mismatch is not explicitly addressed. These limitations are largely practical in nature. Future work may explore hybrid strategies that combine physics-informed trajectory optimization with structured feedback, extensions to systems with explicit actuator constraints or uncertainties, and online or adaptive variants of the framework to enhance robustness and real-time applicability.

\section{Conclusion}

In this work, we have proposed a PINN framework for continuous-time synchronization regulation in networked nonlinear dynamical systems. By jointly parametrizing system trajectories and control inputs within a unified neural representation and explicitly enforcing the governing dynamics, synchronization objectives, and control regularization during training, the proposed approach enables direct regulation of both the synchronization level and the timescale over which coherence emerges.

Using networked Kuramoto oscillators as a representative testbed, we demonstrated that the PINN-based control achieves smooth synchronization without impulsive transients and attains a cumulative control cost that closely approaches the efficiency of analytically motivated frequency-compensation strategies in gradient dynamics, while avoiding the large early-time power peaks characteristic of phase-based feedback. Importantly, the framework remains effective in nongradient and frustrated settings, where conventional baselines lose their theoretical justification.

Beyond the specific numerical examples, the principal contribution of this work is methodological. By embedding macroscopic synchronization requirements directly into a continuous-time, physics-constrained optimization problem, the PINN framework transforms synchronization regulation into a programmable design task. Trade-offs between synchronization accuracy, convergence time, and control expenditure can be explored systematically without prescribing explicit control laws.

The Appendixes~\ref{app:A}--\ref{app:E} further support these results through additional tests, including extensions to limit-cycle oscillators with amplitude dynamics and demonstrations at moderately larger network sizes. While practical challenges remain in computational cost and offline implementation, these results indicate that the framework remains effective across different dynamical regimes when suitable collective observables are available.

Taken together, these results position physics-informed neural control as a promising and versatile tool for the regulation of collective dynamics, offering a unified perspective that bridges dynamical systems theory, optimal control, and machine learning.

\begin{figure*}[!t]
    \centering
    \includegraphics[width=1\linewidth]{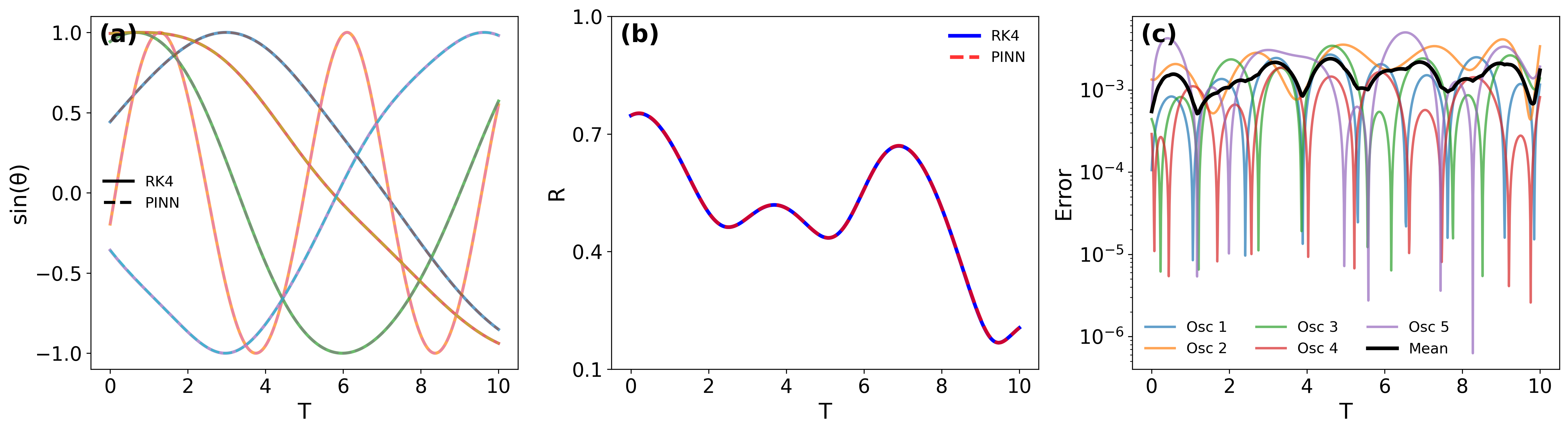}
    \caption{Physics-informed neural network approximation of uncontrolled Kuramoto dynamics in the weak-coupling regime ($K=0.05$). (a) Phase trajectories shown as $\sin\theta_i(t)$, comparing RK4 solutions (solid lines) and PINN predictions (dashed lines).
    (b) Time evolution of the synchronization order parameter $R(t)$.
    (c) Absolute trajectory errors for individual oscillators and their mean value on a logarithmic scale.}
    \label{fig:pinn_uncontrolled_K005}
\end{figure*}
\section*{Data availability}
The code supporting this work is publicly available at Ref.~\cite{ref38}. No additional research data beyond those generated by the code are publicly available.

\appendix
\begin{figure*}[!t]
    \centering
    \includegraphics[width=1\linewidth]{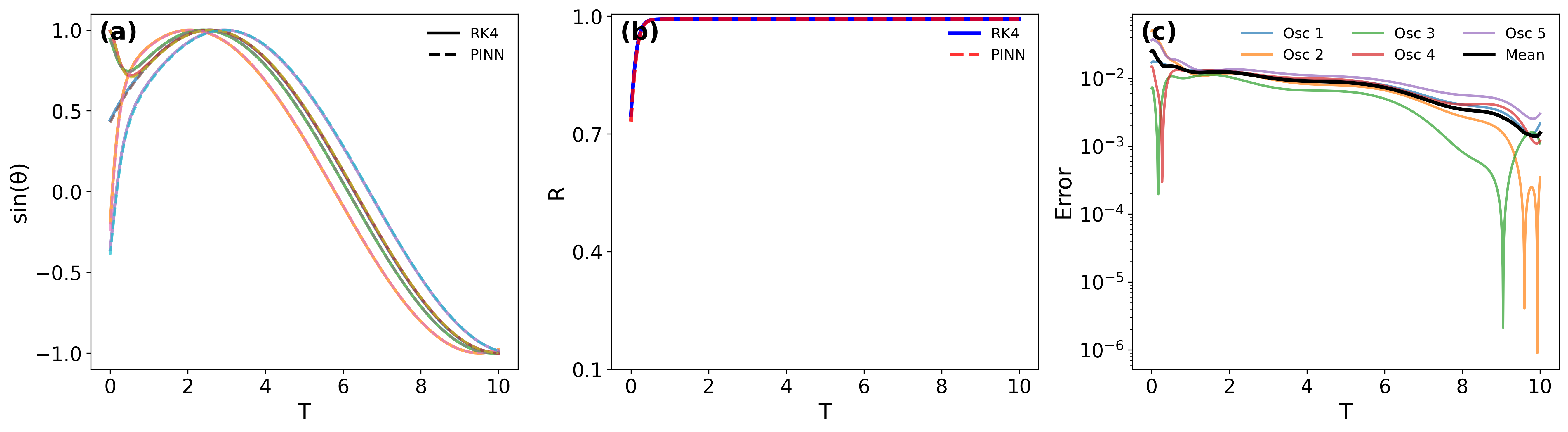}
    \caption{Physics-informed neural network approximation of uncontrolled Kuramoto dynamics in the strong-coupling regime ($K=1.0$). (a) Phase trajectories shown as $\sin\theta_i(t)$.
    (b) Time evolution of the synchronization order parameter $R(t)$.
    (c) Absolute trajectory errors for individual oscillators and their mean value.}
    \label{fig:pinn_uncontrolled_K1}
\end{figure*}

\section{Validation of PINN trajectory representation}
\label{app:A}

Before addressing synchronization control, it is essential to verify that the PINN can faithfully approximate the uncontrolled system dynamics. This validation ensures that subsequent control optimization is carried out on an accurate and physically consistent surrogate of the underlying differential equations, rather than on artifacts introduced by model approximation errors.

We consider the uncontrolled Kuramoto model
\begin{equation}
\dot{\theta}_i(t)
=
\omega_i
+
K \sum_{j=1}^{N} A_{ij}
\sin\!\bigl(\theta_j(t)-\theta_i(t)\bigr),
\end{equation}
where $\theta_i(t)$ denotes the phase of oscillator $i$, $\omega_i$ is its intrinsic frequency, $K$ is the global coupling strength, and $A_{ij}$ represents the adjacency matrix. Throughout this section, no external control is applied, i.e., $u_i(t)\equiv 0$, and the PINN is trained solely by minimizing the physics-informed residual associated with the governing equations, together with the imposed initial conditions.

The residual associated with the uncontrolled dynamics is defined as
\[
r_i(t) = \dot{\theta}_i(t) - \left[ \omega_i + K \sum_{j=1}^N A_{ij} \sin(\theta_j(t) - \theta_i(t))  \right],
\]
and we evaluate its magnitude $\text{Error}(t)=|r(t)|$ along the learned trajectories.

Unless otherwise stated, all results are obtained for a fully connected network of $N=5$ oscillators. 
This small system size is adopted to allow for a detailed diagnostic assessment of trajectory-level approximation accuracy, without obscuring individual dynamics by high-dimensional state representations.  Reference trajectories are computed using a fourth-order Runge-Kutta (RK4) scheme with a sufficiently small time step to guarantee numerical accuracy.

To probe the performance of the PINN across qualitatively distinct dynamical behaviors, two representative coupling regimes are examined. A weak coupling case ($K=0.05$) corresponds to an incoherent regime, while a strong coupling case ($K=1.0$) leads to spontaneous synchronization driven by collective interactions.

In the weak-coupling regime, the PINN accurately reconstructs the continuous-time evolution of individual oscillator phases despite the absence of phase locking. As illustrated in Fig.~\ref{fig:pinn_uncontrolled_K005}, representative trajectories displayed as $\sin\theta_i(t)$ show excellent agreement between PINN predictions and RK4 solutions. At the macroscopic level, the time evolution of the synchronization order parameter $R(t)$ remains low and fluctuating, consistent with incoherent dynamics, and is faithfully reproduced by the PINN. The absolute trajectory errors for individual oscillators, as well as their mean value, remain uniformly small over the entire time interval.

A complementary assessment is provided by the strong-coupling regime $K=1.0$, in which the system undergoes spontaneous synchronization. Figure~\ref{fig:pinn_uncontrolled_K1} demonstrates that the PINN successfully captures both the transient phase-locking process and the long-time synchronized state. The growth and saturation of the order parameter $R(t)$ are accurately reproduced, and the approximation errors remain small and stable despite the presence of strong collective interactions.

These results establish that the PINN provides an accurate representation of uncontrolled Kuramoto dynamics across both incoherent and synchronized regimes. This validation serves as a necessary prerequisite for the control studies presented in the main text and in the following Appendixes.

The PINN-based fitting of the time-series dynamics, along with the corresponding implementation details, has been presented in Ref.~\cite{ref38} in the main text.

\begin{figure*}[!t]
    \centering
    \includegraphics[width=1\linewidth]{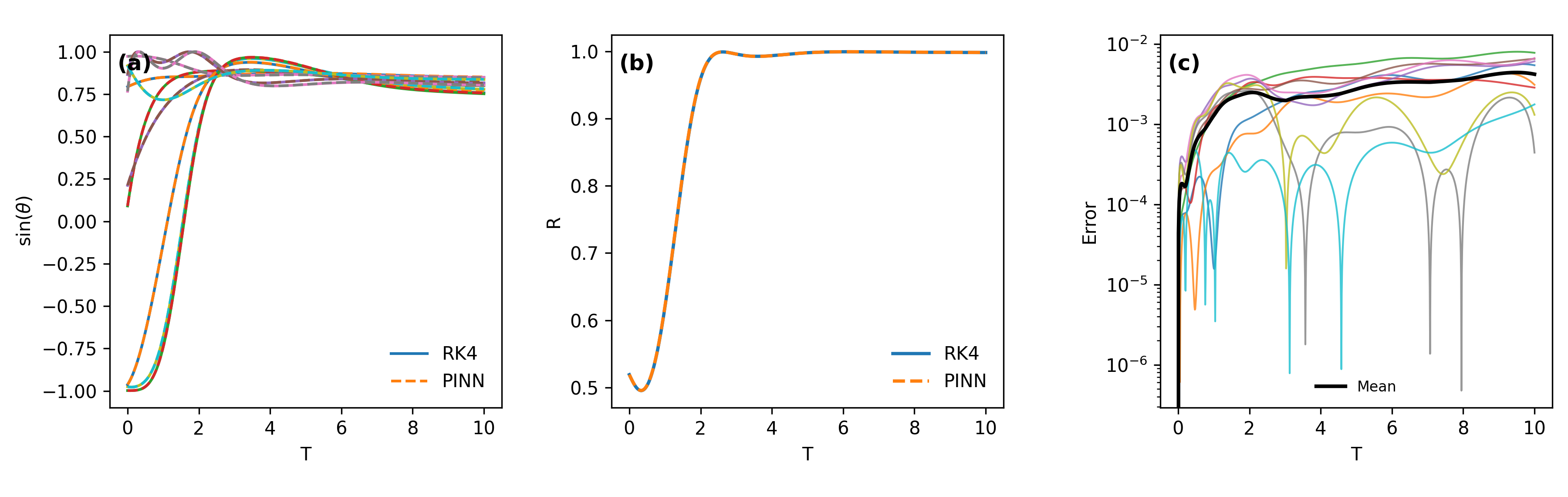}
    \caption{
    Dynamical consistency of PINN-controlled trajectories. (a) Phase trajectories $\sin\theta_i(t)$ comparing RK4 reference solutions (solid lines) and PINN predictions (dashed lines).
    (b) Time evolution of the order parameter $R(t)$.
    (c) Absolute residual $|r_i(t)|$ for individual oscillators and their mean value.
    }
    \label{fig:validation_control}
\end{figure*}

\begin{figure*}[!t]
    \centering
    \includegraphics[width=1\linewidth]{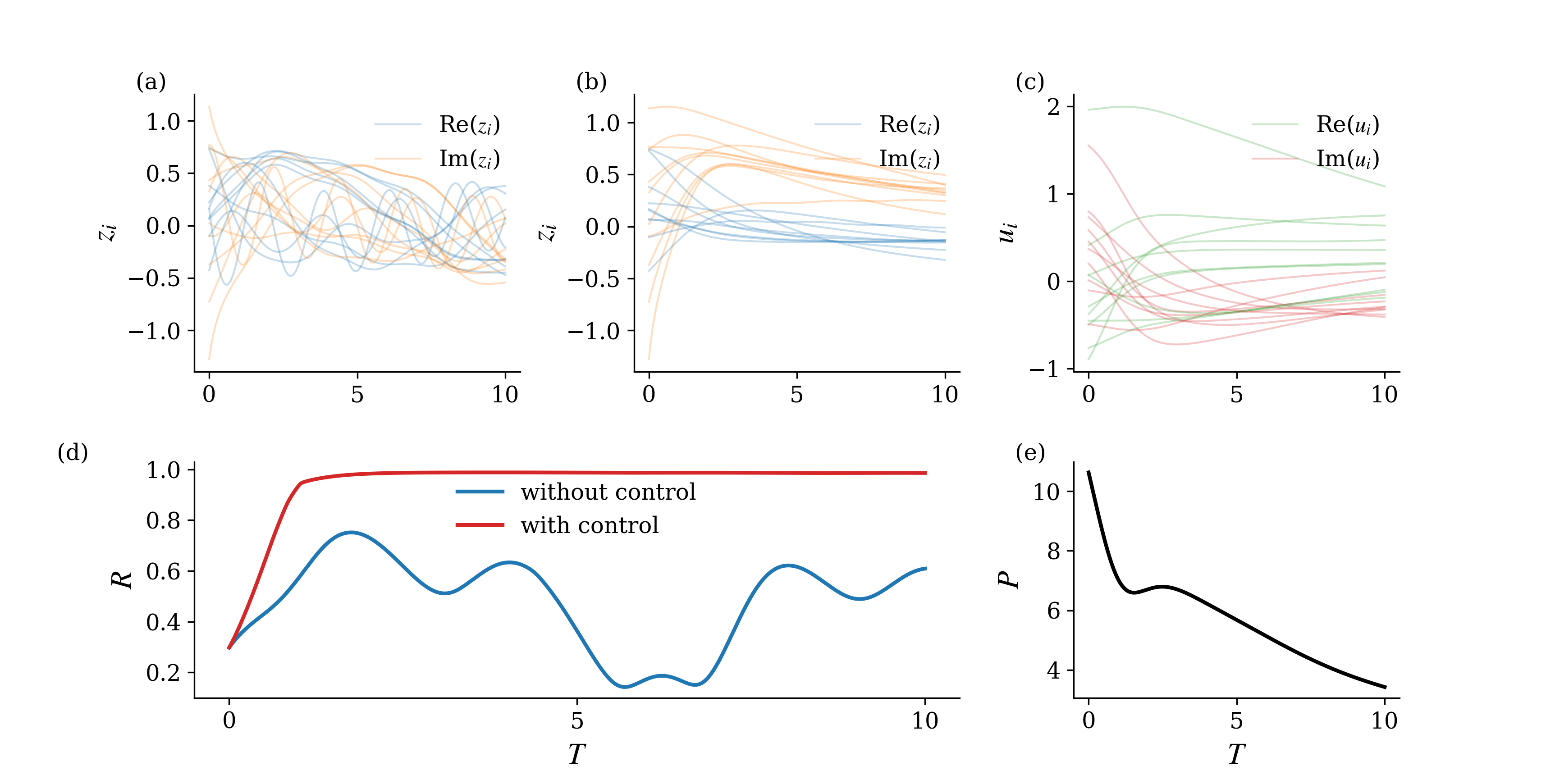}
    \caption{Validation of PINN control for Stuart--Landau dynamics. (a) Complex trajectories without control.
    (b) Complex trajectories under PINN control.
    (c) Learned complex-valued control signals.
    (d) Order parameter $R(t)$ with and without control.
    (e) Instantaneous control power $P(t)$.}
    \label{fig:SL_validation}
\end{figure*}

\section{Dynamical consistency of controlled trajectories}
\label{app:B}

To verify that the learned controlled trajectories remain consistent with the governing dynamics, we evaluate the physics-informed residual of the controlled system. This provides a quantitative diagnostic of whether the PINN-generated trajectories satisfy the underlying differential equations beyond visual agreement.

We consider the same Kuramoto system in the main text, with control inputs $u_i(t)$ learned by the PINN. The residual, extending the definition in Appendix~\ref{app:A} to include the control inputs, associated with the controlled dynamics is defined as
\[
r_i(t) = \dot{\theta}_i(t) - \left[ \omega_i + K \sum_{j=1}^N A_{ij} \sin(\theta_j(t) - \theta_i(t)) + u_i(t) \right],
\]
and we evaluate its magnitude  $\text{Error}(t)=|r(t)|$ along the learned trajectories.

Figure \ref{fig:validation_control} shows a direct comparison between the PINN-controlled trajectories and reference solutions obtained using an RK4 integrator. The phase trajectories [Fig.~\ref{fig:validation_control}(a)] and the corresponding order parameter $R(t)$ [Fig.~\ref{fig:validation_control}(b)] exhibit excellent agreement, indicating that the learned control successfully reproduces the expected synchronization dynamics.

A more stringent assessment is provided by the residual error shown in [Fig.~\ref{fig:validation_control}(c)], where we plot the absolute residual for each oscillator as well as their mean value. The residual remains small throughout the entire time interval, typically on the order of $10^{-3}$--$10^{-2}$, with transient spikes at early times due to initialization. This confirms that the learned controlled trajectories satisfy the governing equations with high accuracy.

These results demonstrate that the PINN does not merely fit trajectories but produces dynamically consistent solutions that respect the controlled Kuramoto dynamics, thereby supporting the validity of the control optimization results reported in the main text.

\begin{figure*}[!t]
    \centering
    \includegraphics[width=0.85\linewidth]{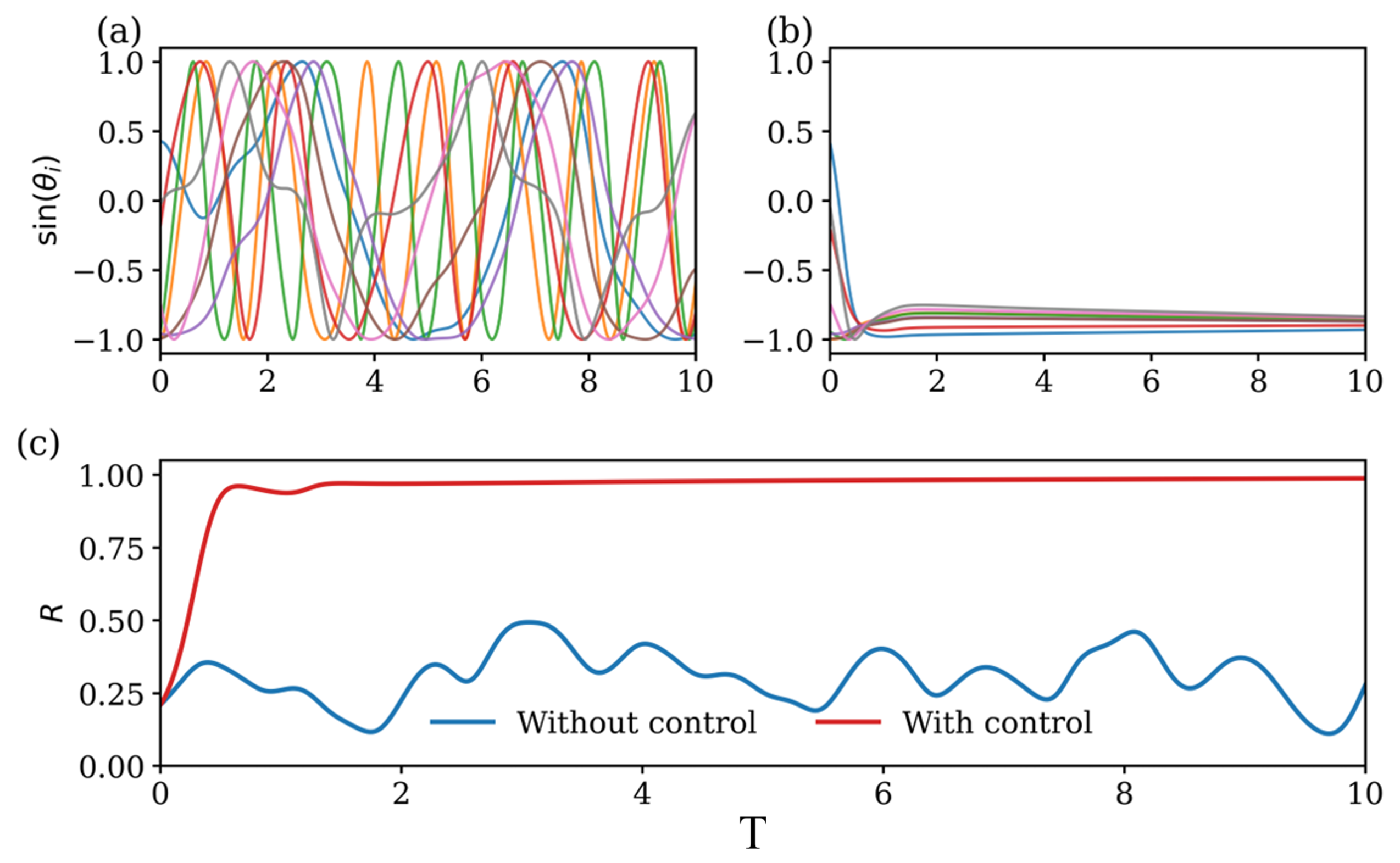}
    \caption{
Time evolution for a Kuramoto network with $N=100$ oscillators. (a) Time series of $\sin\theta_i(t)$ for a representative subset of eight oscillators without control.
(b) Time series of the same oscillators with control.
(c) Corresponding Kuramoto order parameter $R(t)$.
}
    \label{fig:N100_timeseries}
\end{figure*}

\section{Validation of PINN control for Stuart--Landau dynamics}
\label{app:C}
\label{sec:SM_SL_validation}

To demonstrate that the proposed PINN-based control framework is not limited to phase-only oscillator models, we extend the analysis to networks of Stuart--Landau oscillators. This system represents the canonical normal form near a Hopf bifurcation and explicitly incorporates both amplitude and phase degrees of freedom, thereby providing a stringent test of the method's generality.

The dynamics of a network of $N$ diffusively coupled Stuart--Landau oscillators with control are given by
\begin{equation}
\dot{z}_i(t)
=
\bigl(\lambda + i\omega_i - |z_i(t)|^2\bigr) z_i(t)
+
K \sum_{j=1}^{N} A_{ij} \bigl(z_j(t) - z_i(t)\bigr)
+
u_i(t),
\label{eq:SL_control}
\end{equation}
where $z_i(t)\in\mathbb{C}$ denotes the complex amplitude, $\lambda>0$ is the self-excitation parameter, $\omega_i$ is the natural frequency, and $u_i(t)$ is a complex-valued control input learned by the PINN.

Within the PINN framework, both the state and control are represented through their real and imaginary components, allowing the complex-valued dynamics to be enforced via automatic differentiation. Synchronization is quantified using the normalized complex order parameter
\begin{equation*}
R(t)
=
\frac{\bigl|\sum_{i=1}^{N} z_i(t)\bigr|}
{\sum_{i=1}^{N} |z_i(t)|},
\end{equation*}
which simultaneously captures phase coherence and amplitude alignment. This normalized complex order parameter jointly captures phase coherence and amplitude alignment, and reduces to the standard Kuramoto order parameter in the weak amplitude--fluctuation limit. 

Figure~\ref{fig:SL_validation} summarizes the control performance. In the absence of control, the oscillators exhibit dispersed limit-cycle dynamics in the complex plane. Under PINN control, the trajectories progressively collapse onto a coherent manifold, indicating effective regulation of both amplitude and phase. The corresponding order parameter rapidly increases toward unity and remains stable thereafter. The learned control signals are strongest during the transient synchronization stage and decrease substantially once coherence is established, reflecting adaptive control effort.

This example confirms that the PINN-based control strategy naturally extends to systems with coupled amplitude and phase dynamics, without requiring any modification of the underlying framework.

\begin{table}[!htb]
\centering
\caption{Hyperparameters and training configuration used in the main-text experiments.}
\label{tab:hyperparameters}
\begin{tabular}{ll}
\hline
Category & Specification \\
\hline

\textbf{Neural network} & \\
Architecture & MLP (state + control) \\
Hidden layers & 4 \\
Hidden width & 60 \\
Activation & $\tanh$ \\
Input normalization & $\tau = 2t/T - 1$ \\
Control bound $u_{\max}$ & 5 \\
\hline
\textbf{Training} & \\
Optimizer & Adam \\
Learning rate & $3\times10^{-3}$ \\
Training steps & 3000 \\
Collocation points $n_f$ & 2000 \\
IC diagnostic samples $n_b$ & 256 \\
\hline
\textbf{Loss weights} & \\
$w_{\mathrm{phys}}$ & 1000 \\
$w_{R}$ & 500 \\
$w_{u}$ & $10^{-4}$\\
\hline
\textbf{Synchronization target} & \\
Target level $R^\ast$ & 1 \\
Switching time $t^\ast$ & 3 \\
Time window & $[t^\ast, T]$ \\
\hline
\textbf{Evaluation} & \\
Integrator & RK4 \\
Time step $\Delta t$ & 0.01 \\
Control interpolation & Linear \\
\hline
\end{tabular}
\end{table}

\section{Feasibility for a moderately large network ($N=100$)}
\label{app:D}
\label{sec:SM_N100}

To assess the feasibility of the proposed approach beyond small-scale demonstrations, we consider a Kuramoto network consisting of $N=100$ oscillators. The goal of this test is to verify that the method remains trainable and effective for moderately large system sizes, rather than to explore asymptotic scalability.

The natural frequencies are independently drawn from a uniform distribution $\omega_i\sim\mathcal{U}(-2\pi,2\pi)$. The interaction network is fully connected, and the initial phases are sampled from $\theta_i(0)\sim\mathcal{U}(-\pi,\pi)$. The PINN architecture follows the same design principles as in the main text, with two hidden layers of width $200$ for both the state and control networks. In practice, choosing the number of hidden neurons to be on the order of twice the number of oscillators provides sufficient expressive capacity to represent the high-dimensional dynamics and associated control signals.

Representative results are shown in Fig.~\ref{fig:N100_timeseries}. Without control, the oscillators exhibit persistent incoherent dynamics. Once the learned control is applied, the phases rapidly align and remain synchronized. The corresponding order parameter clearly transitions from a low-synchrony regime to a stable synchronized state with $R(t)\approx 1$.

These results demonstrate that the PINN-based control framework remains practically feasible for networks of moderate size, provided that the neural network architecture is chosen with sufficient width to accommodate the increased dimensionality.

\section{Training protocol and hyperparameters}
\label{app:E}

The PINN is used to learn both the phase trajectories $\theta_i(t)$ and the control inputs $u_i(t)$ for a Kuramoto network. The full training protocol and hyperparameter settings used in the main-text experiments are summarized in Table~\ref{tab:hyperparameters}. Unless otherwise stated, the same configuration is employed across different experiments, with only system parameters or synchronization targets varied. The implementation is available at Ref.~\cite{ref38}.

The synchronization objective is defined by a target level $R^\ast=1$ and a target time $t^\ast$, corresponding to complete synchronization in the idealized formulation. In practice, this condition is implemented in a numerical sense: synchronization is considered achieved when the order parameter satisfies $R(t)\ge 0.995$ for all $t\ge t^\ast$, thereby accounting for finite numerical precision and stochastic variability.

The composite loss function includes a regularization term $L_{\mathrm{reg}}$ weighted by $w_{\mathrm{reg}}$ (denoted as $w_u$ in Table~\ref{tab:hyperparameters}), which promotes low-energy control solutions while preserving dynamical consistency and satisfaction of the synchronization constraint. In practice, we find that choosing $w_{\mathrm{reg}}$ in the range $10^{-6}$--$10^{-4}$ yields consistent results. Within this range, the learned trajectories accurately satisfy the governing dynamics, and the qualitative behavior of the control signals as well as their relative comparison with baseline strategies remain robust, although the absolute control cost varies with $w_{\mathrm{reg}}$.

\FloatBarrier

\end{document}